\documentstyle[12pt,epsfig]{article}
\begin{document}
\begin{center}
{\large{\bf QCD Effective Coupling in the Infrared Region}}

\vspace*{5mm}

{G.~Ganbold}\footnote{{\tt ~~ganbold@theor.jinr.ru}}

\vspace*{3mm}

{ Bogoliubov Lab. Theor. Phys., JINR, 141980, Dubna, Russia; \\
Institute of Physics and Technology, 210651, Ulaanbaatar, Mongolia}
\end{center}

\begin{abstract}
We estimate the QCD effective charge $\alpha_s$ in the low-energy region by
exploiting the conventional meson spectrum within a relativistic quantum-field
model based on analytic confinement. The ladder Bethe-Salpeter equation is
solved for the masses of two-quark bound states. We found a new, independent
and specific infrared-finite behavior of  QCD coupling below energy scale 1~GeV.
Particularly, an infrared-fixed point is extracted at $\alpha_s(0)\simeq 0.757$
for confinement scale $\Lambda=345$ MeV.  As an application, we estimate
masses of some intermediate and heavy mesons and obtain results in reasonable
agreement with recent experimental data.
\end{abstract}

\vskip 5mm


\section{Introduction}

The study of QCD behavior at large distances is an active field of research
in particle physics because many interesting and novel behaviors are expected
at low energies below 1 GeV (see, e.g., \cite{pros03,PDG2008}). Understanding
of a number of  phenomena such as quark confinement, hadronization, the
effective coupling and nonvanishing vacuum expectation values, etc. requires
a correct description of hadron dynamics in the infrared (IR) region. However,
the well-established conventional perturbation theory cannot be used effectively
in the IR region and it is required either to supply some additional
phenomenological parameters (e.g., ''effective masses'', anomalous vacuum
averages, etc.), or to use some nonperturbative methods (lattice simulations
\cite{davi08}, power correction \cite{doks96}, string fragmentation \cite{gorc01},
Dyson-Schwinger equations, etc.).  There exists a phenomenological indication
in favor of a smooth transition from short distance to long-distance physics
\cite{doks96}.

One of the fundamental parameters of nature,  the QCD effective coupling
$\alpha_s$ can provide a continuous interpolation between the asymptotical
free state, where perturbation theory works well, and the hadronization
regime,  where nonperturbative techniques must be employed.

QCD predicts the functional form of the energy dependence of $\alpha_s$ on
energy scale $Q$, but its actual value at a given $Q$ must be obtained from
experiment.  This dependence is described theoretically by the renormalization
group equations and measured at relatively high energies \cite{chek03,beth00}.
A self-consistent and physically meaningful prediction of the QCD effective
charge in the IR regime remains one of the actual problems in particle physics.

The present paper is aimed to determine the QCD effective charge in the
low-energy region by exploiting the hadron spectrum. In doing so we extend
our previous investigations \cite{efim02,ganb05,ganb09}, where we provided
new, independent, analytic and numerical estimates on the lowest glueball mass,
conventional meson spectrum and the weak decay constants by using a fixed
(''frozen'') value of  $\alpha_s$. The obtained results were in reasonable
agreement with  experimental evidence.

Below we take into account the dependence of $\alpha_s$ on mass scale $M$
and develop a phenomenological model to describe the IR behavior of $\alpha_s$.
We determine the meson masses by solving the ladder Bethe-Salpeter (BS)
equations for two-quark bound states. The consideration is based on a
relativistic quantum-field model with analytic confinement  (AC) and has a
minimal number of parameters, namely, the confinement scale $\Lambda$ and the
constituent quark masses $m_f,~(f=\{ud,s,c,b\})$.  First, we derive the meson mass
formula and adjust the model parameters by fitting heavy meson masses
($M \geq 2$ GeV). Hereby,  we determine corresponding values of $\alpha_s(M)$
from a smooth interpolation of the newest experimental data on the QCD coupling
constant. Having adjusted model parameters, we estimate $\alpha_s(M)$ in the
low-energy domain by exploiting meson masses below $\sim 1$ GeV. As an
application, we estimate some intermediate and heavy meson masses
($1<M<9.5$ GeV). Finally, we extract a specific IR-finite behavior of the
QCD coupling and conclude briefly recalling the comparison with often quoted
results and recent experimental data.

\section{Effective Coupling of QCD}

The polarization of QCD vacuum causes two opposite effects: the color charge
$g$ is screened by the virtual quark-antiquark pairs and antiscreened by the
polarization of virtual gluons. The competition of these effects results in a
variation of the physical coupling under changes of distance $\sim 1/Q$,
so QCD predicts a dependence $\alpha_s \doteq g^2/(4\pi)=\alpha_s(Q)$.
This dependence is described theoretically by the renormalization group
equations and determined experimentally at relatively high energies
\cite{chek03,beth00}.

\begin{figure}[hb]
\hskip -15mm
\centerline{\includegraphics[width=80mm,height=80mm]{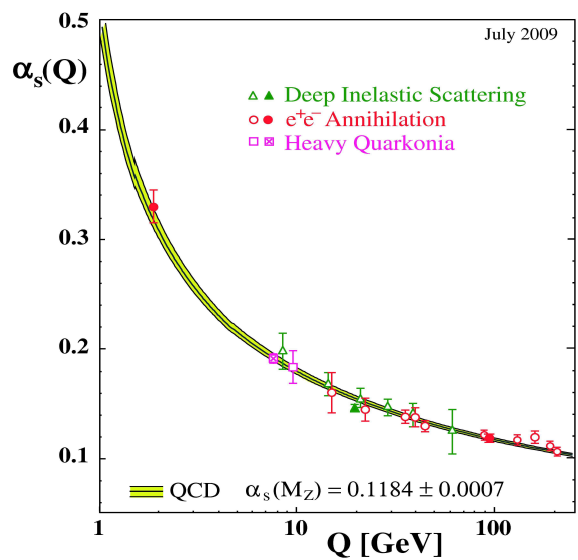}}
\caption{Measurements of $\alpha_s$ as a function of the respective
energy scale $Q$  versa QCD predictions (curves) \cite{beth09}.}
\end{figure}

Nowadays, determinations of $\alpha_s$ remain at the forefront of experimental
studies and tests of QCD. Recent developments on this subject were summarized
in a number of articles \cite{PDG2008,beth09,beth06}.  Summary of the recent
experimental measurements of $\alpha_s$ (Fig. 1) and particular values of
$\alpha_s$ at intermediate energies (Tab. 1) are given by referring to
\cite{beth00,beth09}.

Note that  there are two separate $q^2$ scale regions in which a running coupling
may be considered. The {\it spacelike} region ($q^2 = -Q^2<0$ with relativistic
momentum transfer $Q^2$) is related to scattering  processes while
{\it timelike} domain ($q^2 = M^2>0$, where $M$ is the hadron mass) is often
used for  annihilation and decay processes. The consistent description of QCD
effective coupling $\alpha_s$  in these domains remains the goal of many
studies because only asymptotically the two definitions can be identified but
at low momentum they can be very different (see, e.g. \cite{pros06}).

Particularly, the behavior of one-loop analytic running coupling \cite{nest05}
in timelike and spacelike domains is plotted in Fig. 2.

Many quantities in hadron physics are affected by the IR behavior of the
coupling in different amounts. Nevertheless, the long-distance behavior of
$\alpha_s$ is not well defined, it needs to be more specified
\cite{shir02,nest03,kacz05} and correct description of QCD effective
coupling in the IR regime remains one of the actual problems in particle
physics. Particularly, one of the most precise determinations of $\alpha_s$
near the low-energy region is done by studying $\tau$-lepton decays reporting
central values ranging from 0.318 to 0.344 \cite{bene08,nari09,davi08b}.

\begin{figure}[thb]
\hskip -15mm
\centerline{\includegraphics[width=60mm,height=50mm]{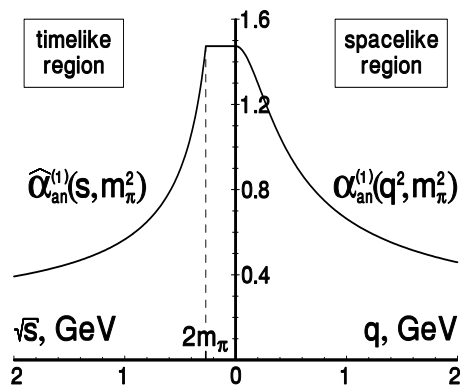}}
\caption{The one-loop massless analytic running coupling in the
spacelike and timelike domains (taken from \cite{nest05}).}
\end{figure}

An attempt to extrapolate the perturbative approach to the long-distance QCD
has been made, it has been suggested that  $\alpha_s$ freezes at a finite and
moderate value \cite{matt94}, and this behavior could be the reason for the
soft transition between short and long distance behaviors.

\begin{table}[ht]
\begin{center}
\begin{tabular}{|c|c|c|c|}
 \hline
Process               & $Q{\mbox{\rm~[GeV]}}$ & $\alpha_s(Q)$ & Ref:  \\
\hline
$\tau$-decays         & 1.78                               & 0.330 $\pm$ 0.014  & \cite{beth09}  \\
$Q\bar{Q}$ states  &  4.1                                 & 0.239 $\pm$ 0.012 & \cite{davi03}  \\
$\Upsilon$ decays &  4.75                               & 0.217 $\pm$ 0.021& \cite{peni98}  \\
$Q\bar{Q}$ states  &  7.5                                 & 0.1923 $\pm$ 0.0024  & \cite{beth09} \\
$\Upsilon$ decays &  9.46                               & 0.184 $\pm$ 0.015 & \cite{beth09}  \\
$e^{+} e^{-}$ jets        &  14.0                               & 0.170 $\pm$ 0.021 & \cite{movi02} \\
\hline
\end{tabular}
\end{center}
\caption{Some measurements of $\alpha_s$ at intermediate energies.}
\end{table}

Different nonperturbative approaches have been proposed to deal with the IR
properties of $\alpha_s$. Particularly, methods, based on gauge-invariant SDE,
concluded that an IR-finite coupling constant may be obtained from first
principles \cite{agui02}.  New solutions for the gluon and ghost SDE have been
obtained with better approximations which led to a new  value for the IR
coupling constant at the origin \cite{bloc02,zwan02}.  Many works within the
lattice simulations have been devoted in recent years to the study of the QCD
running coupling constant either in the perturbative regime \cite{capi99,ster07}
or in the deep IR domain \cite{bouc04}. Note that the results of various
nonperturbative methods for the QCD invariant coupling may differ among
themselves in the IR region due to the specifications of the used methods and
approximations. Particularly, the results obtained by lattice simulations and
SDE methods demonstrate a considerable variety of IR behaviors of
$\alpha_s$.

An extraction of experimental data of $\alpha_s^{exp}(Q^2)$ below 1 GeV
compared with the meson spectrum within analytic perturbation theory has
been performed \cite{bald07} and a summary of data was presented
(see Fig. 2). The earliest attempts to obtain $\alpha_s$ in the IR region were
made in the framework of the quark-antiquark potential models by using the
Wilson loop method \cite{buch80,pete97,bram99,bake96,godf85}. Convenient
interpolation formulas between the large momentum perturbative expression
and a finite IR-fixed point have been used in hadron spectrum studies with
${\alpha_s^0/\pi} \simeq 0.19 - 0.25$ \cite{godf85}.
Within a fully relativistic treatment  it was shown that  a $\rho$-meson mass
much heavier than the $\pi$ mass could be obtained with
${\alpha_s^0/\pi}\simeq 0.265$ \cite{zhan91}
while a similar result within a one-loop analytic coupling method predicted
${\alpha_s^0/\pi}\simeq 0.44$ \cite{bald05}.
A phenomenological hypothesis was adopted that the gluon acquires an effective
dynamical mass $m_g\approx370 MeV$ (at $\Lambda_{QCD}\approx300MeV$)
that resulted in
${\alpha_s^0/\pi}  \simeq 0.26$ \cite{halz93}.
Various event shape in $e^{+}e^{-}$ annihilation can be reproduced with an
averaged value
$\left\langle  {\alpha_s^0/\pi}  \right\rangle  \simeq 0.2$
on interval $\leq 1~GeV$ \cite{doks96}.

\section{Model}

Color confinement in QCD is an attempt to explain the physics phenomenon
that color charged particles are not observed. However, the  reasons for quark
confinement may be somewhat complicated. Particularly, within a quantum-field
model, the quark confinement may be explained as the absence of quark poles
and thresholds in Green's function. Following this idea, the conception of AC
assumes that the QCD vacuum is realized by the self-dual vacuum gluon fields
which are stable versus local quantum fluctuations and related to the
confinement and chiral symmetry breaking \cite{leut81}. This vacuum gluon
field serves as the true minimum of the QCD effective potential \cite{elis85}.
The vacuum of the quark-gluon system has the minimum at the nonzero self-dual
homogenous background field with constant strength. Then, the quark and gluon
propagators in the background gluon field represent entire analytic functions
in Euclidean space \cite{efned95}. In previous papers \cite{ganb08,ganb09}]
we developed relativistic quantum-field models with AC. Similar ideas have
been  realized in infrared confinement by introducing an IR cutoff within a
Nambu-Jona-Lasino model \cite{eber96,volk00}.

The Bethe-Salpeter equation is an important tool for studying the
relativistic two-particle bound states in a field theory framework \cite{bethe51}.
Numerical calculations indicate that the ladder BS equation with a
phenomenological model can give satisfactory results (for a review, see
\cite{robe94}). Particularly, a BS formalism adjusted for QCD was developed
to extract values of $\alpha_s$ below 1 GeV by comparison with known meson
masses \cite{bald07}.

Our purpose is to investigate QCD effective (running) charge in the low-energy
levels by exploiting the spectrum of conventional mesons.  For the spectra of
two-quark bound states we consider a relativistic quantum-field model based
on analytic (or infrared) confinement and solve the  ladder BS equation.

Following previous papers \cite{ganb08,ganb09} we consider a model Lagrangian:
\begin{equation}
{\cal L}=-{1\over 4}\left(  F^A_{\mu\nu} -g f^{ABC}{\cal A}^B_\mu
{\cal A}^C_\nu \right)^2 +\sum_{f}\left( \bar{q}^a_f\left[ \gamma_\alpha
\partial^\alpha-m_f+g\Gamma^\alpha_C {\cal A}^C_\alpha
\right]^{ab}q^b_f \right)\,,
\label{lagrangian}
\end{equation}
where
${\cal A}^C_\alpha$ is the gluon adjoint representation ($\alpha=\{1,...,4\}$);
$F^A_{\mu\nu}=\partial^\mu {\cal A}_\nu^A-\partial^\nu {\cal A}_\mu^A$;
$f^{ABC}$ is the $SU_c(3)$ group structure constant ($\{A,B,C\}=\{1,...,8\})$;
$q_f^a$ is the quark spinor of  flavor $f$ with color $a=\{1,2,3\}$ and mass $m_f$;
$g$ is the coupling strength,
$\Gamma^\alpha_C=i\gamma_\alpha t^C$; and
$t^C$ is the Gell-Mann matrices.

Remember, that within the model the quark and gluon propagators
$\tilde{S}(\hat{p})$ and $\tilde{D}(p)$ in (\ref{lagrangian}) are entire
analytic functions in the Euclidean space.

\subsection{Confinement and Green's Functions}

The effective charge is strongly governed by the detailed dynamics of the
strong interaction and may depend on some of the most fundamental Green's
functions of QCD, such as the gluon and quark propagators \cite{agui09}.
The Green's functions in QCD are tightly connected to confinement and are
ingredients for hadron phenomenology. However, any widely accepted and
rigorous analytic solutions to these  propagators are still missing. One may
encounter difficulties by defining the  explicit quark and gluon propagator
at the confinement scale.  Nowadays,  IR behaviors of  the quark and gluon
propagators are not well-established and need to be more specified
\cite{shir02}.

The matrix elements of hadron processes at large distance are integrated
characteristics of the vertices, quark and gluon propagators and the solution
of the BS equation should not be too sensitive on the details of propagators.
Taking into account the correct global  symmetry properties and their breaking
(also by introducing additional physical parameters) may be more important than
the working out in detail of propagators (e.g., \cite{feld00}). In previous papers
we exploited  simple forms of quark and gluon propagators \cite{ganb09,ganb08}
which were entirely analytic functions in Euclidean space and behaved similarly
to the explicit propagators dictated by AC \cite{efned95}.

Following  \cite{ganb09} we introduce the quark  propagator as follows:
\begin{eqnarray}
\label{q-propag}
\tilde{S}_{m}^{ab}(\hat{p})
=~\delta^{ab} {i\hat{p}+m_f[1\pm\gamma_5~\omega(m_f/\Lambda)]
\over \Lambda m_f}
\exp\left\{-{p^2+m_f^2\over 2\Lambda^2} \right\}\,,
\end{eqnarray}
where $\hat{p}=p_\mu \gamma_\mu$ and $\omega(z)=(1+z^2/4)^{-1}$. The sign
"$\pm$" corresponds to the self- and antiself-dual modes of the background
gluon fields. In (\ref{q-propag}) chiral symmetry breaking is induced by AC.
The interaction of the quark spin with  the background gluon field results in
a singular behavior $\tilde{S}_{\pm}(\hat{p})\sim 1/m_f$ in the massless limit
$m_f\to 0$. This expresses the zero-mode solution (the lowest Landau level)
of the massless Dirac equation in the presence of  an external gluon background
field and generates a nontrivial quark condensate \cite{ganb09} indicating the
broken chiral symmetry as $m_f\to 0$.

Recent theoretical results predict an IR behavior of the gluon propagator.
A gluon propagator  identical to zero at the momentum origin was considered
in \cite{fisc02,lerc02} while another  propagator was of order $1/m_g^2$ [4],
where $m_g$ is the dynamical gluon mass \cite{alle96}. A renormalization
group analysis \cite{gies02} and numerical lattice studies simulating the
gluon propagator are consistent with an IR-finite behavior \cite{lang02}.
We consider a gluon propagator
\begin{eqnarray}
 \tilde{D}_{\mu\nu}^{AB}(p)=\delta^{AB}\delta_{\mu\nu}
{1-\exp\left(-{p^2/\Lambda^2}\right) \over p^2}  = \delta^{AB} \delta_{\mu\nu}
\int\limits_0^{1/\Lambda^2} \!\!\! ds \, e^{-s p^2} \,.
\label{g-propag}
\end{eqnarray}
It represents a modification of gluon propagator defined in \cite{ganb09} and
exhibits an explicit IR-finite behavior $\tilde{D}(0)\sim 1/\Lambda^2$.  For
simplicity $\tilde{D}(p)$ in (\ref{g-propag}) is given in Feynman gauge.

Note that the propagators in Eqs.~(\ref{q-propag}) and (\ref{g-propag}) do not
have any singularities in the finite $p^2$ plane in Euclidean space,  thus
indicating the absence of a single quark (gluon) in the asymptotic space of
states. In fact, an IR parametrization is hidden in the confinement scale
$\Lambda$.

\subsection{Two-quark Bound States}

We allow that the coupling remains weak ($\alpha_s < 1$) in the hadronization
region. Then, the consideration may be restricted within the ladder approximation
sufficient to estimate the meson spectrum with reasonable accuracy. The
leading-order contribution to the two-quark ($q\bar{q}$) bound states is
determined by the partition function
\begin{eqnarray}
&& Z_{q\bar{q}} = \int\!\!\!\int\!\!{\mathcal{D}}\bar{q}{\mathcal{D}} q
\exp\left\{ -(\bar{q} S^{-1}q)+{g^2\over 2} \left\langle
(\bar{q}\Gamma{\cal A} q )
(\bar{q}\Gamma{\cal A} q ) \right\rangle_D   \right\} \,, \nonumber \\
&& \langle  (\bullet) \rangle_D \doteq \int\!\!{\mathcal{D}} {\cal A}
~e^{-{1\over 2}({\cal A} D^{-1}{\cal A})} (\bullet)\,.
\label{pathint}
\end{eqnarray}

Our model has a minimal number of parameters, namely, the scale of confinement
$\Lambda$ and the constituent quark masses ($m_{ud},m_s,m_c,m_b$).

Below we briefly introduce the basic steps entering into our model on the
example of the quark-antiquark bound states \cite{ganb09} defined by
$Z_{q\bar{q}}$ in (\ref{pathint}).

First, we allocate the one-gluon exchange between colored biquark currents
\begin{eqnarray}
{\cal L}_2={g^2\over2}\sum\limits_{f_1f_2} \int\!\!\!\int dx_1dx_2
\left( \bar{q}_{f_1}(x_1)i\gamma_\mu t^A q_{f_1}(x_1)\right)
D_{\mu\nu}^{AB}(x_1,x_2)
\left(\bar{q}_{f_2}(x_2)i\gamma_\nu t^B q_{f_2}(x_2)\right).
\label{L2}
\end{eqnarray}
and isolate the color-singlet combinations.  We perform a Fierz transformation
\begin{eqnarray*}
(i\gamma_{\mu})\delta^{\mu\nu}(i\gamma_{\nu})
=\sum_{J} C_J \cdot O_J~O_J\,,\qquad J=\{S,P,V,A,T\}\,,
\end{eqnarray*}
where $C_J=\{1,1,1/2,-1/2,0\}$ and $O_J=\{I,i\gamma_{5},
i\gamma_{\mu},\gamma_{5}\gamma_{\mu},  i[\gamma_\mu,\gamma_\nu]/2 \}$.
For systems consisting of quarks with different masses it is important to go
to the relative co-ordinates in the center-of-masses system and introduce
the relative masses $\xi_i \doteq m_i /(m_1+m_2)$. Then, introduce a system of
orthonormalized basis functions $\{U_Q(x)\}$,  where $Q=\{n_r,{\em l},\mu\}$
are the radial, orbital and magnetic quantum numbers. Diagonalize  ${\cal L}_2$
on basis $\{U_{Q}(x)\}$ and use a Gaussian path-integral representation for
the exponential
$$
e^{{\cal L}_2} \doteq
e^{{g^2\over2}\sum\limits_{{\cal N}}({\cal J}_{{\cal N}}^2)}
=\left\langle  e^{g(B_{{\cal N}}{\cal J}_{{\cal N}})}
\right\rangle_B \,, \qquad \left\langle (\bullet) \right\rangle_B \doteq
 \int \prod\limits_N {\mathcal{D}}B_N ~e^{-{1\over2}(B_{{\cal N}}^2)}
 (\bullet) \,,  \qquad \left\langle 1 \right\rangle_B=1
$$
by introducing a colorless biquark current ${\cal J}_{{\cal N}}$ and auxiliary
meson fields $B_{\cal N}$ with ${\cal N}=\{Q,J,f_1,f_2\}$.  Then
$$
Z_{q\bar{q}} = \left\langle
\int\!\!\!\int\!\!{\mathcal{D}}\bar{q}{\mathcal{D}} q
\exp\left\{ -(\bar{q} S^{-1}q)+ g(B_{{\cal N}}{\cal J}_{{\cal N}})\right\}
\right\rangle_B \,.
$$
By taking explicit path integration over quark variables we obtain
$$
Z_{q\bar{q}} \rightarrow Z= \left\langle \exp\left\{ {\rm Tr}
\ln\left[1+g(B_{{\cal N}}V_{{\cal N}})S\right]\right\}  \right\rangle_B \,,
$$
where $V_{\cal N}(x,y)$ is a vertex function.

Introduce a {\sl hadronization ansatz} and this will identify $B_{\cal N}(x)$
with meson fields carrying quantum  numbers ${\cal N}$. We isolate all
quadratic field configurations ($\sim B^2_{\cal N})$ in the ''kinetic'' term
and rewrite the partition function for mesons \cite{ganb09}:
\begin{eqnarray}
 Z  = \int {\prod\limits_{\cal N}  {\mathcal{D}}B_{\cal N}  \,\exp
 \left\{ { - \frac{1}{2}  \sum\limits_{{\cal NN'}}  (B_{\cal N} \,
 [\delta^{{\cal NN'}} + \alpha_s\lambda_{\cal NN'} ]\,B_{{\cal N'}} )
 - W_{res} [B_{\cal N} ]} \right\}}  \,,
\end{eqnarray}
where the interaction between mesons is described by the residual part
$ W_{res} [B_{\cal N}] \sim 0(B_{\cal N}^3)$.

The leading-order term of the polarization operator is
\begin{eqnarray}
\alpha_s\lambda_{{\cal NN'}}(z)\doteq  \int\!\!\!\int\!\! dx dy \,
U_{\cal N}(x) \, \alpha_s\lambda_{JJ'}(z,x,y) \, U_{{\cal N'}}(y)\,,
\end{eqnarray}
and the Fourier transform of its kernel reads
\begin{eqnarray}
&& \alpha_s\lambda_{{JJ'}}(p,x,y) \doteq \alpha_s\int\!\! dz \, e^{ipz}
\lambda_{JJ'}(z,x,y)   \nonumber\\
&& =  {4 g^2 \sqrt{C_J\,C_{J'}}\over 9} \sqrt{D(x)D(y)}
\int\!\! {d^4 k\over(2\pi)^4}~e^{-ik(x-y)}
{\rm Tr}\left[O_J \tilde{S}_{m_1}\left(\hat{k}+\xi_1\hat{p}
\right) O_{J'} \tilde{S}_{m_2}\left(\hat{k}-\xi_2\hat{p}\right) \right]\,,
\label{Bethe1}
\end{eqnarray}
where ${\rm Tr}\doteq {\rm Tr}_c{\rm Tr}_\gamma \sum_{\pm}$; ${\rm Tr}_c$
and ${\rm Tr}_\gamma$ are traces taken on color and spinor indices,
correspondingly, while $\sum_{\pm}$ implies the sum over self-dual and
antiself-dual modes.

We diagonalize the polarization kernel on the orthonormal basis $\{U_{\cal N}\}$:
$$
\int\!\!\!\int\!\! dx dy \, U_{{\cal N}}(x) \lambda_{{JJ'}}(p,x,y) U_{{\cal N}'}(y)
=\delta^{{\cal NN}'}~\lambda_{{\cal N}}(-p^2)
$$
that is equivalent to the solution of the corresponding ladder BS equation.
We rewrite
\begin{eqnarray}
\label{Bethe2}
 \lambda_{{\cal N}}(-p^2) &=& {8\, C_J \over 3\pi^3}
\int \!\! {d^4 k} \left| V_J(k) \right|^2 \Pi_{\cal N}(k,p) \,, \\
V_J(k) &\doteq& \int\!\! d^4x\, U_J(x) \sqrt{D(x)}\, e^{-ikx} \,, \nonumber \\
\Pi_{\cal N}(k,p) &\doteq& {1\over 24} {\rm Tr}\left[O_J \tilde{S}_{m_1}
\left(\hat{k}+\xi_1\hat{p} \right) O_{J'} \tilde{S}_{m_2}
\left(\hat{k}-\xi_2\hat{p}\right) \right]\,,\nonumber
\end{eqnarray}
where $V_J(k)$ is a vertex and $\Pi_{\cal N}(k,p)$ is the kernel of the
polarization operator.

In relativistic quantum-field theory  a stable bound state of $n$ massive
particles shows up as a pole in the S-matrix with a center of mass energy.
Accordingly, the physical  mass of the meson may be derived from the equation
\begin{eqnarray}
1+\alpha_s\lambda_{{\cal N}}(M_{{\cal N}}^2)=0\,, \qquad -p^2=M_{\cal N}^2 \,.
\label{Bethe3}
\end{eqnarray}

Then, with a renormalization
\begin{eqnarray}
\label{renormed}
&& (B_{\cal N}[1+\alpha_s\lambda_{\cal N}(-p^2)]B_{\cal N})
=(B_{\cal N}[1+\alpha_s\lambda_{{\cal N}}(M_{{\cal N}}^2)
+\alpha_s\dot\lambda_{\cal N}(M_{\cal N}^2)[p^2+M_{\cal N}^2] B_{\cal N})  \\
&& = (B_R [p^2+M_{\cal N}^2]B_R)\,, \quad \dot\lambda_{\cal N}(z) \doteq
{d\lambda_{\cal N}(z) \over dz} \,, \quad B_R(x) \doteq \sqrt{\alpha_s
\dot\lambda_{\cal N}(M_{\cal N}^2)} \cdot B_{\cal N}(x)
\nonumber
\end{eqnarray}
the partition function  takes the conventional form
\begin{eqnarray}
 Z  = \int \! {\mathcal{D}}B_R  \, \exp  \left\{
 - \frac{1}{2} \left(B_R \left[  p^2+M_{\cal N}^2 \right]  B_R \right)
 - W_{res} [B_R ] \right\}  \,.
\end{eqnarray}

\subsection{Conventional  Meson Spectrum and Running Coupling}

We use the meson mass $M$ as the appropriate characteristic parameter,
so the coupling $\hat{\alpha}_s(M)$ is defined in a timelike domain. On the
other hand, most of known data on $\alpha_s(Q)$ are possible in the spacelike
region. The continuation of the invariant charge from the spacelike to the
timelike region (and vice versa) was elaborated by making use of the integral
relationships between the QCD running coupling in Euclidean and Minkowskian
domains (see, e.g. \cite{milt97,nest03}).

Below we consider the most established sectors of hadron spectroscopy,  the
pseudoscalar ${\mathbf{P}} (0^{-+})$ and vector ${\mathbf{V}} (1^{--})$ mesons.

The dependence of meson masses on $\hat{\alpha}_s$ and other parameters
is defined by Eq. (\ref{Bethe3}).  Note that the polarization kernel
$\lambda_{{\cal N}}(-p^2)$ is natively obtained real and symmetric that
allows us to find a simple  variational solution to this problem. Choosing
a trial Gaussian function for the ground state \cite{ganb09}
\begin{eqnarray}
\label{testf}
U(x) = {2a\over \pi} \exp\left\{-{a \Lambda^2 x^2}\right\}\,,
\qquad \Lambda^4 \!\! \int\!\! d^4x \left| U(x)\right|^2 =1\,, \qquad a>0\,.
\label{trial1}
\end{eqnarray}
we obtain a variational form of Eq. (\ref{Bethe3}) for meson masses
as follows:
\begin{eqnarray}
\label{ground}
1 &=& -\hat{\alpha}_s(M_J)\cdot \lambda_J(\Lambda,M_J,m_1,m_2)  \\
&=& {8 \hat{\alpha}_s C_J  \over 3\pi^2 (m_1/\Lambda)(m_2/\Lambda)}
\cdot\exp\left\{ { M^2_J - (m_1+m_2)^2 \over 2\Lambda^2 } (\xi_1^2+\xi_2^2)
 \right\}                                \nonumber \\
&& \cdot \max\limits_{0<c<2} \left[ c\,(2-c)^2 \right] \! \int\!\!\!\!\int
\limits_{0}^{1} \! {du~ dw \over \sqrt{(1/u-1)(1/w-1)}\, Q^2} \exp\left\{
- {M^2_J \, (\xi_1-\xi_2)^2 \over 4\Lambda^2 \, Q} \right\}         \nonumber\\
&& \cdot \left\{ {2 \, \rho_J \over Q} + {M^2_J \over \Lambda^2}
\left[\xi_1 \xi_2 + {(\xi_1-\xi_2)^2\over 2Q}\left(1-{\rho_J\over 2Q} \right)
\right]  + {m_1 m_2 \over \Lambda^2}
 \left[
 1+\chi_J \, \omega\left(m_1\over\Lambda\right) \omega\left(m_1\over\Lambda\right)
\right]  \right\}  \nonumber \,,
\end{eqnarray}
where $Q\doteq 1+c(u+w)$,  $\rho_J=\{1,1/2\}$ and $\chi_J=\{1,-1\}$ for
$J=\{P,V\}$.

Further we exploit Eq. (\ref{ground}) in different ways, by solving either for
$\hat{\alpha}_s$ at given masses, or for $M_J$ at known values of coupling.
In doing so, we adjust the model parameters by fitting available experimental
data.

Note that any physical observable must be independent of the particular scheme
and mass by definition, but in (\ref{ground}) we obtain  $\alpha_s$ depending on
scaled masses $\{M_J/\Lambda$, $m_1/\Lambda$ and  $m_2/\Lambda \}$, where
$\Lambda$  is the scale of confinement. This kind of scale
dependence is most pronounced in leading-order QCD and often used to test
and specify uncertainties of theoretical calculations for physical observables.
Conventionally, the central value of $\alpha_s(\mu)$ is determined or taken for
$\mu$ equaling the typical energy of the underlying scattering reaction.
There is no common agreement of how to fix the choice of scales.
Particularly, in \cite{ganb09} we fixed the parameter $\Lambda$ by fitting
light meson weak decay constants.

Below we solve Eq. (\ref{ground}) for different values of confinement scale.
As a particular case, first we choose $\Lambda_1=345$ MeV.

1) We can extract intermediate values of $\alpha_s(M_V)$ in interval $2 - 10$
GeV from a smooth interpolation of known data from Table~1. Particularly,
\begin{equation}
\left\{
\begin{tabular}{l}
$\hat{\alpha}_s(9460)=0.1817$\,, \\
$\hat{\alpha}_s(3097)=0.2619$\,, \\
$\hat{\alpha}_s(2112)=0.3074$\,, \\
$\hat{\alpha}_s(2010)=0.3138$\,. \\
\end{tabular}
\right.
\label{interpolate}
\end{equation}
Hereafter,  masses are given in units of $MeV$.

Then, we adjust the constituent quark masses $\{m_{ud},m_s,m_c,m_b \} $  by
solving a set of equations:
\begin{equation}
\left\{
\begin{tabular}{l}
$ 1+\hat{\alpha}_s(9460) \cdot \lambda_V(\Lambda_1,9460,m_b,m_b)=0$\,, \\
$ 1+\hat{\alpha}_s(3097) \cdot \lambda_V(\Lambda_1,3097,m_c,m_c)=0$\,, \\
$ 1+\hat{\alpha}_s(2112) \cdot \lambda_V(\Lambda_1,2112,m_s,m_c)=0$\,, \\
$ 1+\hat{\alpha}_s(2010) \cdot \lambda_V(\Lambda_1,2010,m_{ud},m_c)=0$ \\
\end{tabular}
\right.
\label{fit1}
\end{equation}
with known masses of mesons $\Upsilon(9460)$, $J/\Psi(3097)$, $D^*_s(2112)$
and $D^*(2010)$.  We fix a particular set of model parameters as follows:
\begin{eqnarray}
\Lambda=\Lambda_1=345 {\mbox{\rm ~MeV}} \,,  \qquad
m_{ud}=192.56{\mbox{\rm ~MeV}} \,,              \nonumber\\
m_s=293.45{\mbox{\rm ~MeV}}  \,,\qquad m_c=1447.59{\mbox{\rm ~MeV}}  \,,
\qquad m_b=4692.51{\mbox{\rm ~MeV}} \,.
\label{parameters}
\end{eqnarray}

2)  Having fixed the model parameters,  we solve an inverse problem, to find
$\alpha_s$ values in the region below 1 GeV as follows:
\begin{equation}
\left\{
\begin{tabular}{l}
$ \hat{\alpha}_s(138)=-\lambda^{-1}_P(\Lambda_1,138,m_{ud},m_{ud})=0.7131   $\,, \\
$ \hat{\alpha}_s(495)=-\lambda^{-1}_P(\Lambda_1,495,m_{ud},m_s)=0.6086    $\,, \\
$ \hat{\alpha}_s(770)=-\lambda^{-1}_V(\Lambda_1,770,m_{ud},m_{ud})=0.4390 $\,,\\
$ \hat{\alpha}_s(892)=-\lambda^{-1}_V(\Lambda_1,892,m_{ud},m_s)=0.4214 $\,. \\
\end{tabular}
\right.
\label{fit2}
\end{equation}

In Fig. 3 we plot our low-energy estimates (\ref{fit2}) in comparison with
the three-loop analytic coupling, its perturbative counterpart (both normalized
at the Z-boson mass), and the massive one-loop analytic coupling \cite{bald07}.

\begin{figure}[thb]
\centerline{\includegraphics[width=140mm,height=80mm]{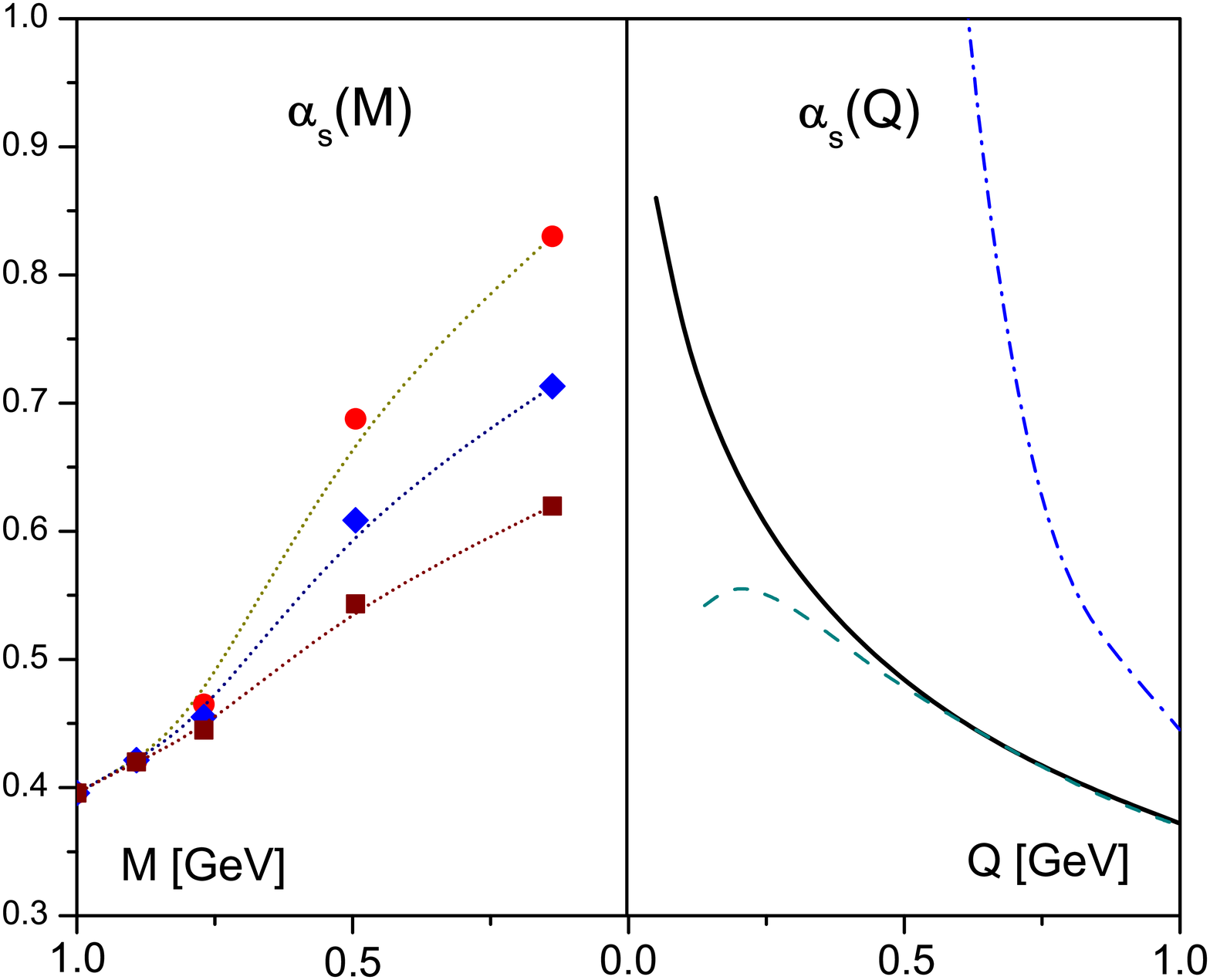}}
\caption{ Our estimates of $\hat{\alpha}_s(M)$ in the low-energy region at different 
values of confinement scale (red dots for $\Lambda=330$ MeV; blue diamonds for
$\Lambda=345$ MeV and black squares for $\Lambda=360$ MeV) compared
with the three-loop analytic coupling $\alpha_s(Q)$ (solid curve), its perturbative
counterpart (dot-dashed curve), and the massive one-loop analytic coupling
(dashed curve) (see Ref. \cite{bald07}).}
\end{figure}

3) As an application, with particular choice of parameters (\ref{parameters})
we calculate masses of other mesons: $D(1870)$, $D_s(1970)$, $\eta_c(2980)$,
$B(5279)$, $B^{*}(5325)$, $B_s(5370)$,  $B_c(6286)$ and $\eta_b(9389)$.
Hereby, the corresponding $\alpha_s(M)$ are extracted from Fig. 1.

Our estimates of meson masses along experimental data \cite{PDG2008} are
shown in Table 2. The relative error of our estimate does not exceed $3.5\%$
percent in a wide range of mass.

\begin{table}[ht]
\begin{center}
\begin{tabular}{|c|c||c|c||c|c||c|c|}
 \hline
$J^{PC}=0^{-+}$ &$M_{\mathbf{P}}$&$J^{PC}=0^{-+}$ & $M_{\mathbf{P}}$
&$J^{PC}=1^{--}$ &$M_{\mathbf{V}}$&$J^{PC}=1^{--}$ & $M_{\mathbf{V}}$\\
 \hline
 $\pi(138)$      & 138   & $\eta_c(2980)$& 3039& $\rho(770)$    & 770
 &$D^*_s(2112)$  & 2112\\
 $K(495)$       & 495   & $B(5279)$     & 5339& $\omega(782)$  & 785
 &$J/\Psi(3097)$ & 3097\\
 $\eta(547)$    & 547  & $B_s(5370)$   & 5439& $K^*(892)$     & 892
 &$B^*(5325)$    & 5357\\
 $D(1870)$     & 1941& $B_c(6286)$   & 6489& $\Phi(1019)$   & 1022
 &$\Upsilon(9460)$&9460\\
 $D_s(1970)$ & 2039& $\eta_b(9389)$& 9442& $D^*(2010)$    & 2010
 &               &     \\
\hline
\end{tabular}
\end{center}
\caption{Masses $M$ of conventional mesons (in units of {\rm MeV})
corresponding to effective coupling $\hat{\alpha}_s(M)$ determined by
Eq. (\ref{ground}) at $\Lambda=345$ MeV.}
\end{table}

4) To check the sensibility of the obtained results on the confinement scale
value  we recalculated steps 1-3 for $\Lambda=330$ MeV and
$\Lambda=360$ MeV. We revealed that the estimated meson masses shown
in Table 2 do not change considerably (less than $0.5\%$ percent). The variation
of $\hat{\alpha}_s$ under changes of $\Lambda$ is shown in Fig. 3.

5) We perform global evaluation of $\hat{\alpha}_s(M)$ at the mass scale of
conventional mesons (shown in Table 2) by using the formula
$$
\hat{\alpha}_s(M_J) = - 1/ \lambda _J(M_J, \Lambda, m_1, m_2)
$$
and we plot the resulting curves at different $\Lambda$ in Fig. 5 in comparison
with recent low- and high-energy data of $\alpha_s(Q)$  \cite{bald07}.

\subsection{IR-finite Behavior of Effective Coupling}

The possibility that the QCD coupling constant features an IR-finite behavior
has been extensively studied in recent years (e.g., \cite{brod04,agui04}).
There are theoretical arguments in favor of a nontrivial IR-fixed point,
particularly,  the analytical coupling freezes at the value of $4\pi/\beta_0$
within one-loop approximation \cite{shir97}. The phenomenological evidence
for $\alpha_s$ finite in the IR region is much more numerous.

We note that the agreement of our estimates of $\hat{\alpha}_s(M)$ with other
predictions (e.g., \cite{beth00,pros06}) turns out to be reasonable from
2 GeV down to the 1 GeV scale.  Below this scale, different  behaviors
of $\alpha_s(M)$ may be expected as $M$ approaches zero.

Below we consider the IR-fixed point $\hat{\alpha}_s^0 \doteq \hat{\alpha}_s(0)$
by evaluating Eq. (\ref{ground}) for $M_P=0$ and $m_1=m_2=m$:
\begin{eqnarray}
\hat{\alpha}_s^0 \!&=&\! {3\pi^2 m^2 \over 8 \Lambda^2}  e^{\mu^2}
\left\{ \max\limits_{0<c<2} \left[ c(2-c)^2\right] \int\!\!\int\limits_{0}^{1}
{du \, dw \over \sqrt{(1/u-1)(1/w-1)}(1+c(u+w))^2} \right. \nonumber\\
&&
\left. \left[{2\over (1+c(u+w))^2} + {\mu^2}(1+\omega^2(\mu))
\right] \right\}^{-1}\,.
\label{alpha0}
\end{eqnarray}
The dependence of $\hat{\alpha}_s^0$ on $\mu \doteq m/\Lambda$ is plotted
in Fig. 4.

\begin{figure}[thb]
\hskip -15mm
\centerline{\includegraphics[width=80mm,height=80mm]{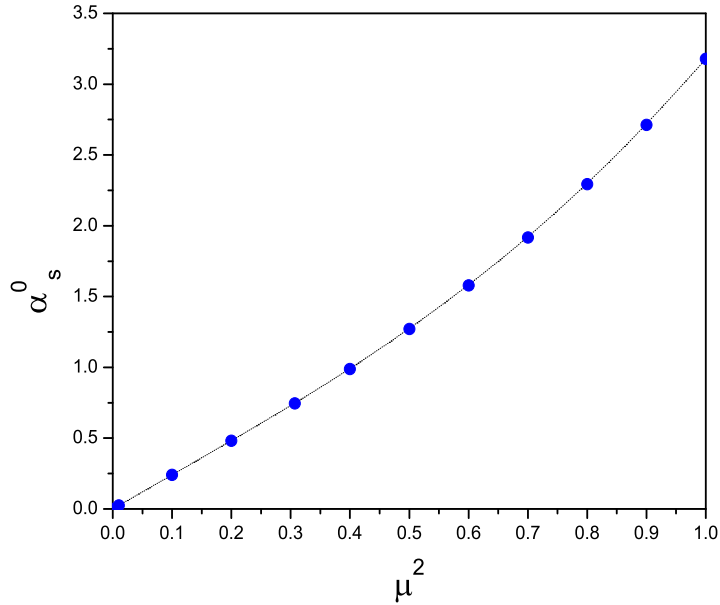}}
\caption{Dependence of IR fixed point $\alpha_s^{0}$ on the scaled
quark mass $\mu^2=(m/\Lambda)^2$  at fixed confinement scale
$\Lambda=345$ MeV.}
\end{figure}

Note that a value of $\hat{\alpha}_s^0$ of order ~$2$ or larger would be
definitely out
of line with many other phenomena, such as nonrelativistic potentials for
charmonium \cite{bada00} and analytic perturbation theory \cite{shir97}.
Obviously, this constraint implies an upper limit to the value of constituent
quark mass: $\mu^2 < 0.8$ or $m< 0.9\, \Lambda$.

Since we are searching the IR-fixed point, it is reasonable to choose the
lightest quark mass.  Particularly, for $m=m_{ud}=192.56 ~\mbox{\rm MeV}$
and  $\Lambda=345~ \mbox{\rm MeV}$ we obtain
\begin{eqnarray}
\hat{\alpha}_s^0=0.757 \,, \qquad \mbox{or} \qquad {\hat{\alpha}_s^0/\pi}=0.241\,.
\label{alphaMyIR}
\end{eqnarray}

To compare our result with known data on $\alpha_s(Q)$ we exploit the
integral relationships between the QCD running coupling in Euclidean and
Minkowskian domains. Particularly, there exists a relation \cite{nest03}
\begin{eqnarray}
\alpha_s(q^2)=q^2\int\limits_{0}^{\infty} { ds \over (s+q^2)^2} \, \hat{\alpha}_s(s)
\label{spacetimelike}
\end{eqnarray}
valid for the case of massless pion. By substituting $s=t\,q^2$ into
(\ref{spacetimelike}) one rewrites
\begin{eqnarray}
\alpha_s(q^2)=\int\limits_{0}^{\infty} { dt \over (1+t)^2} \,  \hat{\alpha}_s(t\,q^2) \,.
\end{eqnarray}
Then, for $q^2\to 0$ we obtain
\begin{eqnarray}
\alpha_s(0)=\hat{\alpha}_s(0) \int\limits_{0}^{\infty} {dt \over (1+t)^2}
= \hat{\alpha}_s(0)\cdot 1\,.
\end{eqnarray}

Therefore, we may conclude that our result (\ref{alphaMyIR}) is in reasonable
agreement with often-quoted estimates
\begin{equation}
\left\{
\begin{tabular}{l}
${\alpha_s^0/\pi} \simeq 0.19 - 0.25$ \qquad  \cite{godf85}\,, \\
${\alpha_s^0/\pi}\simeq 0.265$ \qquad \qquad ~~\cite{zhan91}\,, \\
${\alpha_s^0/\pi}  \simeq 0.26$ \qquad \qquad ~~~\cite{halz93}\,, \\
$\left\langle  {\alpha_s^0/\pi}  \right\rangle_{1\,GeV}  \simeq 0.2$
\qquad ~~~\cite{doks96}
\end{tabular}
\right.
\label{alpha0IR}
\end{equation}
and phenomenological evidences \cite{bald05,bald07}. The obtained IR-fixed
value of the coupling constant is moderate, it depends on the mass of constituent
quark ($u,d$), so one can insert  this value into perturbative expressions to be
compatible with the experimental data.

\begin{figure}[thb]
\centerline{
\includegraphics[width=80mm,height=60mm]{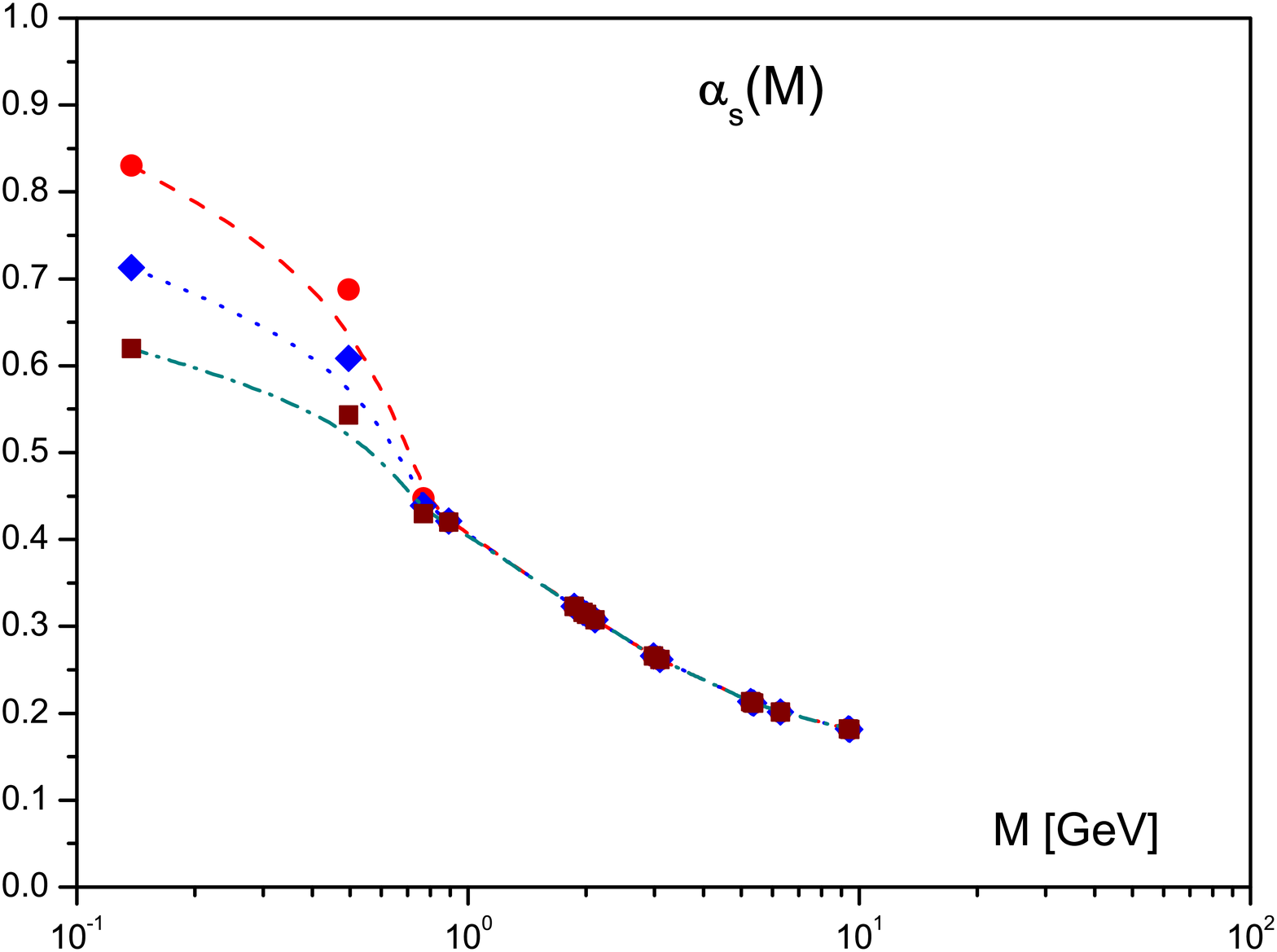}
\includegraphics[width=80mm,height=60mm]{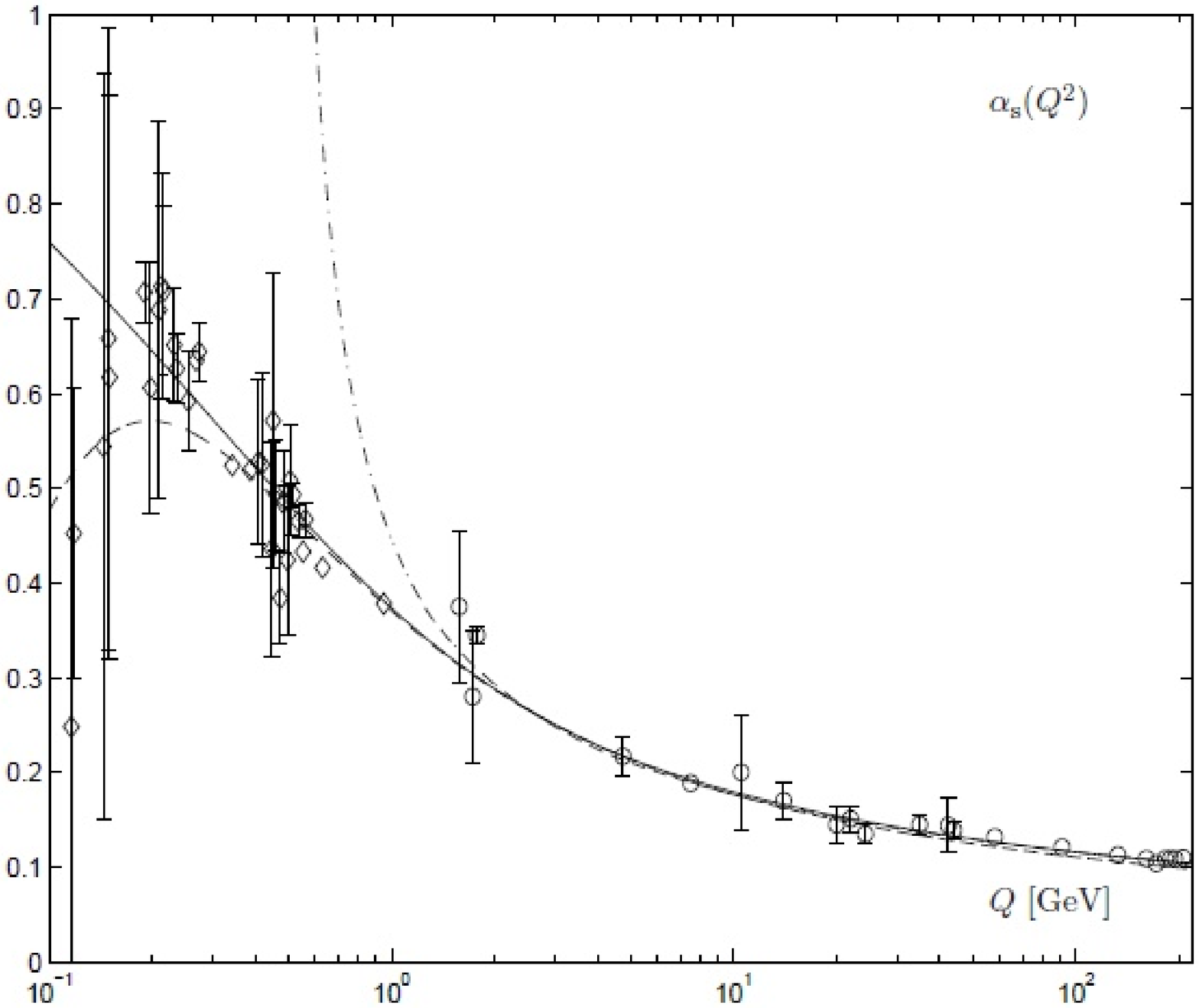}
}
\caption{ Summary of estimates of $\hat{\alpha}_s(M)$ in interval from 0 to 10
GeV at different values of confinement scale. In the left panel
$\Lambda=330$ MeV (red dots), $\Lambda=345$ MeV (blue diamonds) and
$\Lambda=360$ MeV (black squares) compared with $\alpha_s(Q)$ (in the
right panel) defined in low-energy (open diamonds) and high-energy (open circles)
experiments. Also shown are the three-loop analytic coupling (solid curve), its
perturbative counterpart (dot-dashed curve) both normalized at the Z-boson
mass, and the massive one-loop analytic coupling (dashed curve)
(for details see Ref. \cite{bald07}).}
\end{figure}

By interpolating smoothly obtained results in (\ref{alphaMyIR}),  (\ref{fit2})
and (\ref{interpolate}) into the intermediate-energy region we define $\hat{\alpha}_s$
on a wide interval $0.14 - 9.5$ GeV.  Some particular cases of the dependence
$\alpha_s$ on mass scale $M$ at different model parameters are plotted
in Fig. 5.

It is important to stress that we do not aim to obtain the behavior of the
coupling constant at all scales. At moderate $M^2=-p^2$ we obtain $\alpha_s$
in coincidence with the QCD predictions. However, at large mass scale
(above 10 GeV) $\hat{\alpha}_s$  decreases much faster than expected by QCD
prediction. The reason is the use of confined propagators in the form of entire
functions, Eqs.~(\ref{q-propag}) and (\ref{g-propag}). Then, the convolution of
entire functions leads to a rapid decreasing (or a rapid growth in Minkowski
space) of physical matrix elements once the hadron masses and energies of
the reaction have been fixed. Consequently, the numerical results become
sensitive to changes of model parameters at large masses and energies.

\section{Conclusion}

To conclude, we provide an estimate of QCD effective charge in the low-energy
region (below 1 GeV)  by exploiting the conventional meson spectrum within a
relativistic quantum-field model based on analytic (or infrared) confinement.
The new results obtained in the previous section are summarized in
Figs.~3-5 and Table~2.

We demonstrate that global properties of the low-energy phenomena such as
QCD running coupling and conventional meson spectrum may be explained
reasonably in the framework of a simple relativistic quantum-field model of
quark-gluon interaction based on analytic (or, infrared) confinement. Our
guess about the symmetry structure of the quark-gluon interaction in the
confinement region has been tested and the use of simple forms of
propagators has resulted in quantitatively reasonable estimates.

Despite its pure model origin, the approximations used, and questions about the
very definition of the coupling in the IR region, our approach demonstrates a
new, independent and specific IR-finite behavior of  QCD coupling and we
extract a particular IR-fixed point at $\hat{\alpha}_s(0)\simeq 0.757$ for confinement
scale $\Lambda=345$ MeV. As an application,  we performed estimates on
intermediate and heavy meson masses and the result was in reasonable
agreement with experimental data.  Our estimates may be improved further by
using iterative schemes, but the aim is to obtain a qualitative understanding of
QCD effective coupling in the IR region.

The suggested model in its simple form is far from real QCD but we conclude
that the analytic confinement conception combined with BS method may provide
us with a rather satisfactory correlated understanding of  low and
intermediate-energy phenomena from few hundreds MeV to few GeV.

Note that  further improvements of measurements of $\alpha_s$ will be difficult
while it is unlikely that QCD perturbation theory will considerably improve
existing predictions. Therefore, further developments of theoretical
predictions within nonperturbative methods and reapplication of improved
models may have successes in this field.

\vskip 2mm

The author thanks M.A.~Ivanov, E.~Klempt and  A.V.~Nesterenko for useful
discussions and valuable remarks.


\end{document}